\documentclass[aps,showpacs,nofootinbib,twocolumn]{revtex4}

\usepackage{graphicx}
\usepackage{bm}
\usepackage{amsmath}
\usepackage{amssymb}
\usepackage{hyperref}

\newcommand{\xbj}{x}
\newcommand{\zh}{z_h}
\newcommand{\nslash}{\kern 0.2 em n\kern -0.50em /}
\newcommand{\kslash}{\kern 0.2 em k\kern -0.45em /}
\newcommand{\lslash}{\kern 0.2 em l\kern -0.50em /}
\newcommand{\pslash}{\kern 0.2 em p\kern -0.50em /}
\newcommand{\Sslash}{\kern 0.2 em S\kern -0.50em /}
\newcommand{\Pslash}{\kern 0.2 em P\kern -0.50em /}
\newcommand{\Dslash}{\kern 0.2 em D\kern -0.65em /\kern 0.15em}

\newcommand{\bp}{\boldsymbol{p}_T}
\newcommand{\bP}{\boldsymbol{P}_T}
\newcommand{\bk}{\boldsymbol{k}_T}

\newcommand{\eps}{\epsilon}
\newcommand{\ssh}{\!\!\!/}
\newcommand{\Tr}{\operatorname*{Tr}\nolimits}

\newcommand{\ph}{\phi_h}

\begin{document}

\title{Single-spin asymmetries in electroproduction of pions on the longitudinally polarized nucleon targets}

\author{Zhun Lu}\email{zhunlu@seu.edu.cn}\affiliation{Department of Physics, Southeast University, Nanjing
211189, China}

\begin{abstract}
We study the single-spin asymmetries of pions produced in semi-inclusive deep-inelastic scattering on the longitudinally polarized nucleon targets.
We particularly consider the effects of the twist-3 transverse-momentum dependent distribution functions $f_L^\perp$ and $h_L$ to the asymmetry.
We calculate the asymmetric moment $A_{UL}^{\sin\phi_h}$ for $\pi^+$, $\pi^-$ and $\pi^0$ produced off the proton target at HERMES and compare the results with the HERMES data.
We also present the prediction of the same asymmetries for different pions at the kinematics of CLAS 5.5 GeV on a proton target, as well as at COMPASS on a deuteron target.

\end{abstract}

\pacs{12.39.-x, 13.60.-r, 13.88.+e}

\maketitle

\section{Introduction}

Understanding the origins of single-spin asymmetries in high-energy processes is one of the main goals in QCD and hadronic physics~\cite{bdr,D'Alesio:2007jt,Barone:2010ef,Boer:2011fh}.
The longitudinal target spin asymmetry (LTSA)~\cite{Airapetian:2005jc,hermes00,hermes01,hermes03}, as the first unambiguous single-spin phenomenon measured in
Semi-inclusive deep-inelastic scattering (SIDIS), has attracted a lot of theoretical attentions.
Experimentally, the LTSA is usually measured by using the target longitudinally polarized with respect to the incoming beam direction, which makes the interpretation of the asymmetry more complicated.
This is because in this case the measured asymmetry contains
contributions from both the transverse and longitudinal
polarization components with respect to the virtual
photon direction.
The small but non-vanishing transverse polarization component, although suppressed by a factor $1/Q$ compared to the longitudinal component, can contribute to the measured asymmetry through the collins~\cite{Collins:1993npb} or Sivers~\cite{Sivers:1991prd} effects, which are leading-twist observables.
This feature, especially the former one, has been considered
in several estimates~\cite{Boglione2000,Korotkov2001,
Efremov2001,Ma2002,Efremov2003a,Efremov2003b}
in terms of the transversity distribution and
Collins fragmentation functions~\cite{Collins:1993npb}.
Apart from this, There are also competing contributions that originate from the various convolutions of the twist-3 quark distribution functions or fragmentation functions with their twist-2 counterparts, due to the presence of the large longitudinal component.
Therefore, the asymmetry from those contributions, usually denoted by $A_{UL}^{\sin\phi_h}$ where $\phi_h$ is the azimuthal angle of the final hadron around the photon direction, is the purely twist-3 results, compared to the leading-twist ones.

In this paper, we will study the LTSA based upon the contributions from the subleading-twist effects.
Although those effects encode important twist-3 dynamics of the nucleon, they are barely studied in phenomenology and have not been fully considered in the previous theoretical analyses.
At the twist-3 level, there are four individual contributions~\cite{Bacchetta:0611265} that can give rise to the LTSA.
Two of them are from the the twist-3 fragmentation functions, while another two are from the twist-3 distribution functions.
In this work, we will focus on the effects of the twist-3 quark distributions, that is, the distributions $f_L^\perp(x,\bm k_T^2)$ and $h_L(x,\bm k_T^2)$.
The former one is a T-odd transverse momentum dependent (TMD) distribution and can be viewed as the twist-3 analogy of the Sivers function; it generates the LTSA through the coupling with the unpolarized fragmentation function $D_1$.
The later one is a chiral-odd distribution, therefore its contribution to the LTSA involves another chiral-odd fragmentation function, the Collins function $H_1^\perp$~\cite{Collins:1993npb}.
In our study we will consider both the $f_L^\perp D_1$ term~\cite{Metz:2004epja22} and the $h_L H_1^\perp$ term~\cite{Boglione2000}.
Currently the knowledge on the twist-3 distributions is very limited.
In order to quantitatively predict the LTSA in SIDIS contributed by the subleading-twist effect, we will employ the spectator-diquark models to calculate the TMD distributions $f_L^\perp$ and $h_L$ for the $u$ and $d$ valence quarks inside the proton.
As the asymmetry we calculate, the $A_{UL}^{\sin\phi_h}$, corresponds to the case in which the nucleon is longitudinally polarized with respect to the virtual photon momentum, it is not easy to compare our result directly  with the experiments in which the longitudinal spin of the target is originally along the beam direction.
Fortunately, there are available data $A_{UL}^{\sin\phi_h}$ for $\pi^+$ and $\pi^-$ production from the HERMES collaboration, which are extracted from the  original data by subtracting the contributions from the Sivers effect and the Collins effect, using the data on the
transversely polarized target~\cite{hermes05}.
The extraction is valid up to corrections of order $\sin\theta_{\gamma^*}$, with $\theta_{\gamma^*}$ the polar angle between the incoming beam direction and the virtual photon direction.
New data on $A_{UL}^{\sin\phi_h}$ are also coming from the CLAS at the Jefferson Lab.
These make the comparison between theoretical estimates and experimental measurements possible.

The remained content of the paper is organized as follows.
In Section II, we calculate the twist-3 TMD distributions $f_L^\perp$ and $h_L$ using the spectator-diquark models, and we will use two different models to obtain two sets of TMD distributions for comparison.
In section III, we estimate the LTSA contributed by the $f_L^\perp D_1$ and $h_L H_1^\perp$ terms at the HERMES kinematics, and compare the results with the HERMES data.
We will also give the prediction on the LTSA at CLAS and COMPASS.
We summarize the paper in Section IV.

\section{Calculation of distributions $f_L^\perp$ and $h_L$ in spectator-diquark models}
\label{formulation}

\begin{figure}
  \includegraphics[width=0.8\columnwidth]{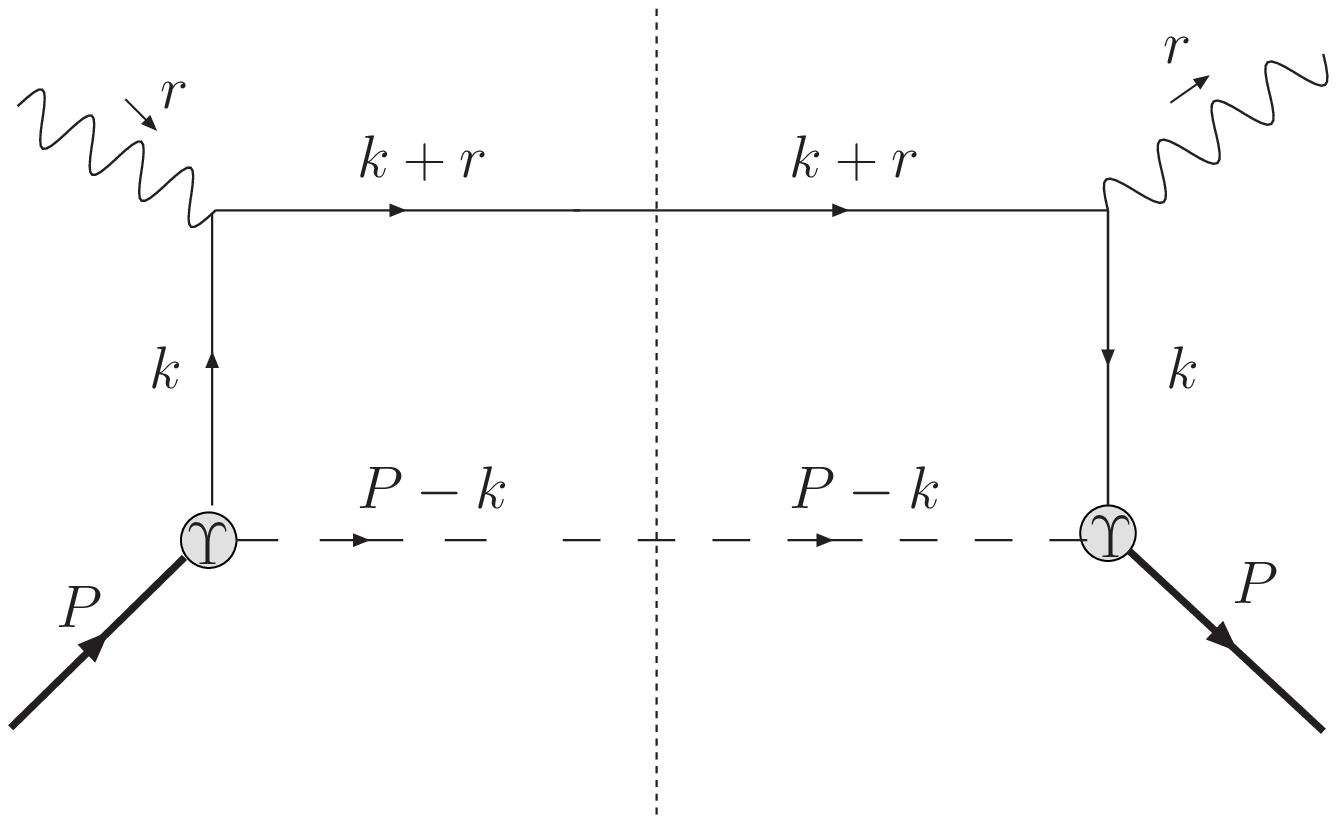}\\
  \includegraphics[width=0.8\columnwidth]{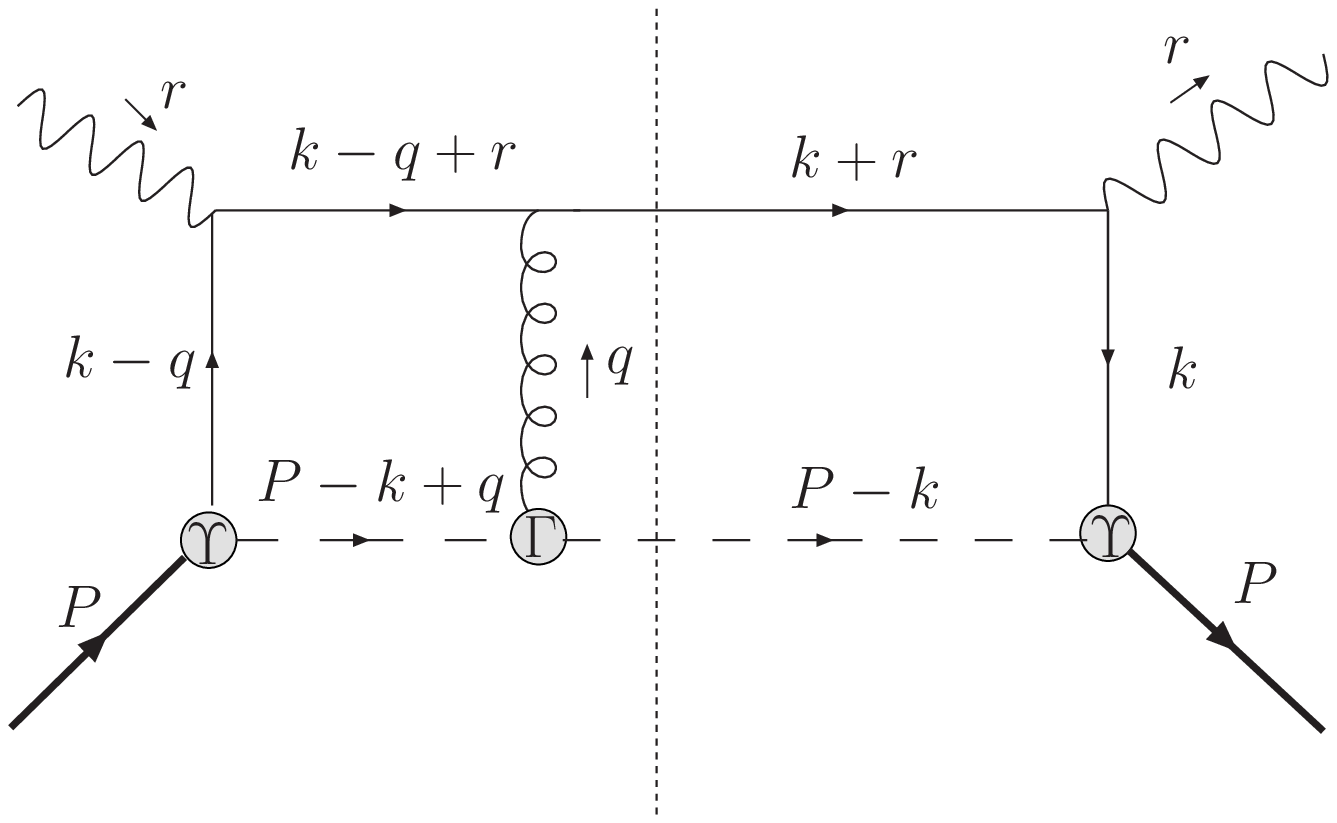}
 \caption{Cut diagrams for the spectator model calculation at tree level (upper) and one-loop level (lower). The dashed lines denote the spectator-diquarks that can be scalar diquarks or axial-vector diquarks.}
 \label{fig:diagrams}
\end{figure}

In this section, we present our calculation on the twist-3 TMD distributions $f_L^\perp(x,\bm k_T^2)$ and $h_L(x,\bm k_T^2)$ for the $u$ and $d$ valence quarks.
The TMD $f_L^\perp(x,\bm k_T^2)$ has been calculated in a scalar diquark model~\cite{Metz:2004epja22,Lu:2012plb}, while $h_L(x,\bm k_T^2)$ has been calculated
by the spectator model~\cite{Jakob:1997npa} and the bag
model~\cite{Avakian:2010prd81}.
Although $h_L(x,\bm k_T^2)$ has the integrated version $h_L(x)$~\cite{Boglione2000}, here we trace its $k_T$-dependence in order to calculate its contribution to $A_{UL}^{\sin\phi_h}$.

The gauge-invariant quark-quark correlator for the longitudinally polarized nucleon can be expressed as
\begin{align}
\Phi^{[+]}(x,\bm k_T)&=\int {d\xi^- d^2\xi_T\over (2\pi)^3}e^{ik\cdot\xi}
\langle P S_L|\bar{\psi}_j(0)\mathcal{L}[0^-,\infty^-]\nonumber\\
& \times \mathcal{L}[\bm 0_T,\bm \xi_T]\mathcal{L}[\infty^-,\xi^-]\psi_i(\xi)|P S_L\rangle\,,
\label{Phi}
\end{align}
where $[+]$, corresponding to the SIDIS process, denotes that the gauge-link appearing in $\Phi$ is future-pointing;
$k$ and $P$ are the momenta of the struck quark and the target nucleon, respectively.
The TMD distributions $h_L$ and $f_L^\perp$ may be obtained from the correlator via the following traces:
\begin{align}
S_L\frac{M}{P^+} h_L(x,\bm k_T^2)   & =
\frac{1}{2}\Tr[\Phi^{[+]}i\sigma^{+-}\gamma_5],\label{eq:hltr}\\
S_L\frac{\eps_{T}^{\alpha\rho} k_{T \rho}^{}}{P^+} \,
  f_L^{\perp}(x,\bm{k}_{T}^{2})& =
-\frac{1}{2}\Tr[\Phi^{[+]}\gamma^{\alpha}]. \label{phitr2}
\end{align}

The correlator (\ref{Phi}) can be calculated by the spectator models~\cite{Jakob:1997npa,Bacchetta:plb578,Gamberg:2007wm,Bacchetta:2008prd}.
Here we consider the contributions of both the scalar diquark and the axial-vector diquark, and use the dipolar form factor for the nucleon-quark-diquark couplings.
The corresponding diagrams used the in the calculation are shown in Fig.~\ref{fig:diagrams}.
In the lowest-order expansion of the gauge-link, which is equivalent to set  $\mathcal{L}=1$,
we apply the diagram in the upper panel of Fig.~\ref{fig:diagrams} to obtain the correlator contributed by the scalar diquark component:
\begin{align}
\Phi^{(0)}_s(x,\bk)&\equiv \frac{N_s^2(1-x)^2}{16 \pi^3 P^+}\frac{\left[ (k\ssh +m)(P\ssh +M)\gamma_5 S\ssh (k\ssh +m)\right]}{(\bk^2+L_s^2)^4}, \label{lophis}
\end{align}
and by the axial-vector diquark component:
\begin{eqnarray}
 \Phi^{(0)}_{v}(x,\bk)
&\equiv& \frac{N_v^2(1-x)^3}{32\pi^3 P^+}d_{\mu\nu}(P-k)\nonumber\\
&\times& \frac{\left[(k\ssh +m)\gamma^{\mu}(M-P\ssh )\gamma_5 S\ssh \gamma^{\nu} (k\ssh+m)\right]}{(\bk^2+L_v^2)^4}, \label{lophiv}
\end{eqnarray}
where $N_s$ and $N_v$ are the normalization constant, $d_{\mu\nu}$ is the polarization sum (the propagator) of the axial-vector diquark, and $L_X^2$ ($X=s$ or $v$) has the form
\begin{align}
L_X^2=(1-x)\Lambda_{X}^2 +x M_{X}^2-x(1-x)M^2.
\end{align}
with $\Lambda_X$ the cutoff parameters for the quark momentum, and $M_X$ the mass for the diquarks.

\begin{figure*}
  \includegraphics[width=0.9\columnwidth]{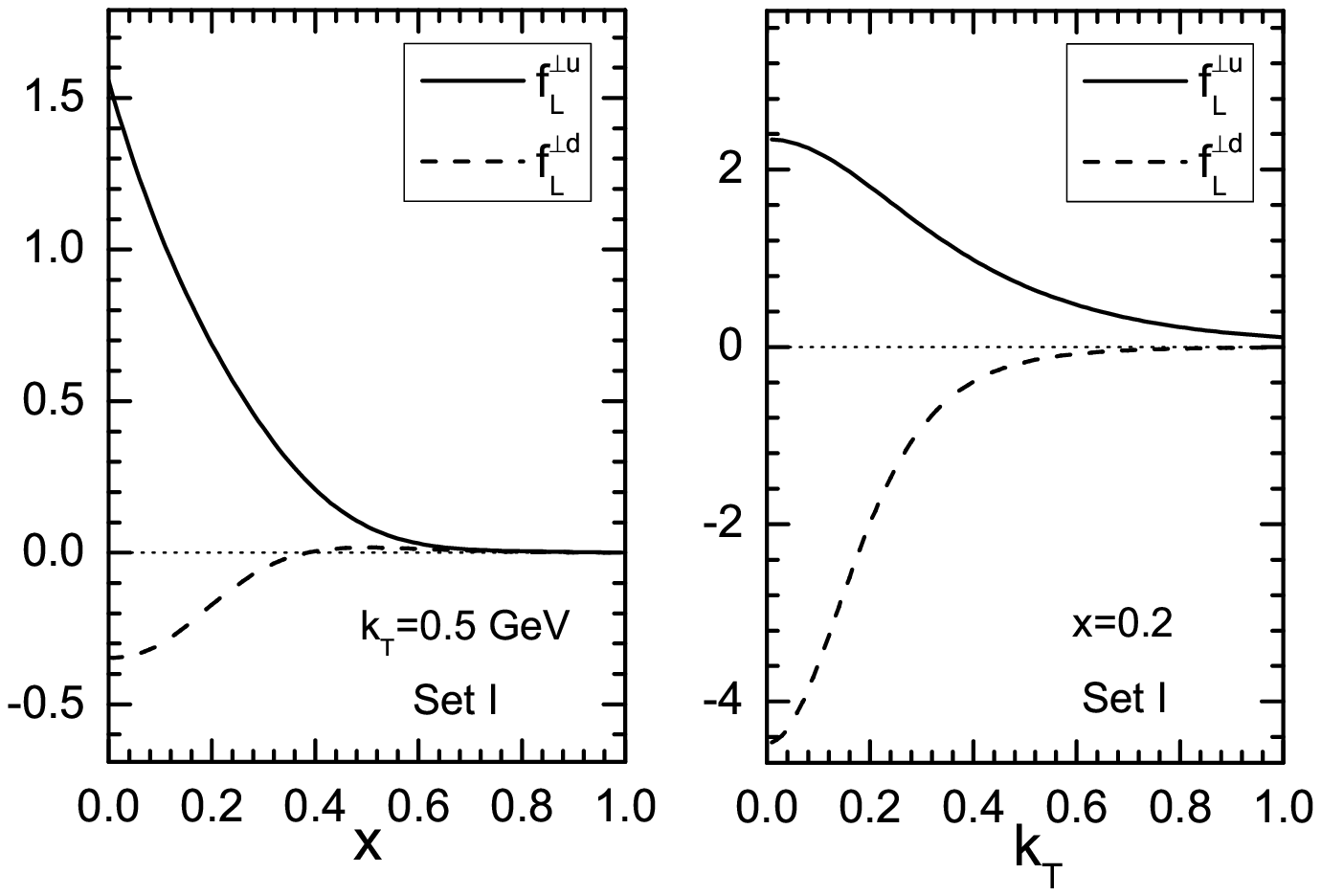}~~~~~~
    \includegraphics[width=0.9\columnwidth]{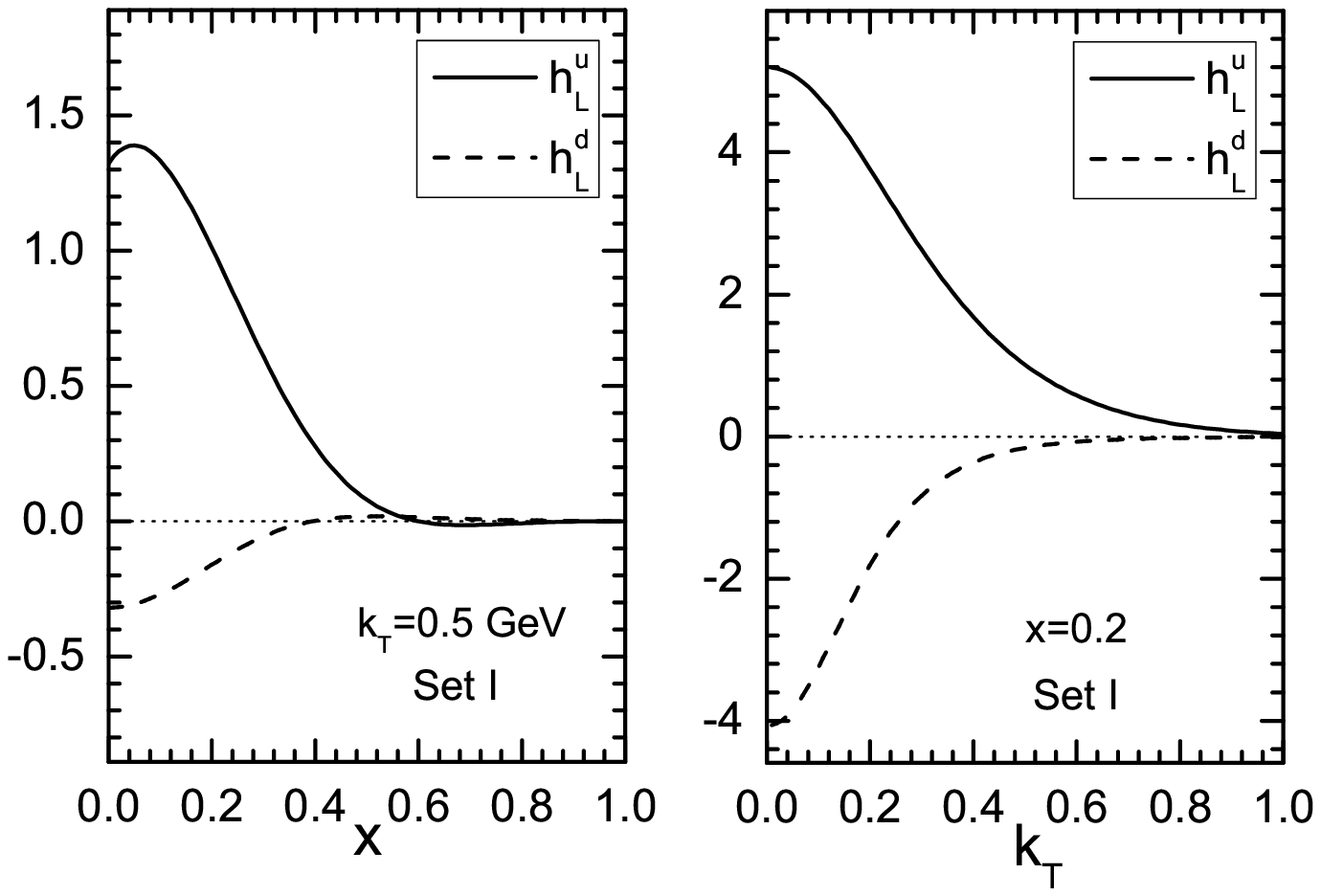}
 \caption{The TMD distributions $f_L^\perp (x,\bm k_T^2)$ (left panel)  and $h_L(x,\bm k_T^2)$ (right panel) for $u$ and $d$ quarks in Set I, calculated from the spectator model in Ref.~\cite{Bacchetta:2008prd}.}
 \label{fig:hlflset1}
\end{figure*}

To calculate the T-odd TMD distribution $f_L^\perp$, one has to consider the nontrivial effect of the gange-link~\cite{Brodsky:2002plb,jy02,Collins:2002plb}, that is, the final-state interaction between the struck quark and the spectator.
Following Refs.~\cite{Bacchetta:plb578,Bacchetta:2008prd}, we perform the calculations initially with Abelian gluons and generalize the result to QCD at the end.
In our calculation we expand the gauge-link to one-loop order, as shown by the diagram in the lower panel of Fig.~\ref{fig:diagrams}.
We note that in Refs.~\cite{Gamberg:2009uk,Gamberg:2010xi}, an eikonal model including a whole ladder of
non-perturbative gluon exchanges was used to describe the final state interactions relevant for T-odd TMDs.
After some algebra we obtain the expressions for the correlator contributed by the scalar diquark and the arxial-diquark component at this order:
\begin{align}
  \Phi_s^{(1)}
(x,\bm k_T)
&\equiv
-i e_q N_{s}^2  { (1-x)^3\over 64\pi^3 (P^+)^2}{-i\Gamma_s^+\over (\bm{k}_T^2+L_s^2)^2}\nonumber \\
 & \hspace{-1cm} \times \int {d^2 \bm q_T\over (2\pi)^2}
{ \left[(\kslash -q\ssh+m) (\Pslash+M)\gamma_5\Sslash(\kslash +m)\right]
\over \bm q_{T}^2  ((\bm{k}_T-\bm{q}_T)^2+L_s^2)^2}
 , \label{phis1}\\
 \Phi^{(1)}_{v}
(x,\bm k_T)
&\equiv
-i e_q N_v^2  { (1-x)^2\over 128\pi^3 (P^+)^2}{1\over (\bm{k}_T^2+L_v^2)^2}\nonumber\\
&\times\int {d^2 \bm q_T\over (2\pi)^2} \,
 d_{\rho\alpha}(P-k)\, (-i\Gamma^{+,\alpha\beta}) \nonumber\\
 &\times d_{\sigma\beta}(P-k+q) \nonumber\\
&\times{ \left[(\kslash -q\ssh+m) \gamma^\sigma(M-\Pslash)\gamma_5\Sslash\gamma^\rho (\kslash +m)\right]
\over \bm q_T^2  ((\bm{k}_T-\bm{q}_T)^2+L_v^2)^2},
\end{align}
where $q^+=0$ is understood, $e_q$ is the charge for the quarks, $d_{\sigma\beta}$ is polarization sum of the vector diuqark which will be determined later, and
$\Gamma_s^\mu $ or $\Gamma_v^{\mu,\alpha\beta}$ is
the vertex between the gluon and the scalar diquark or the axial-vector diquark:
\begin{align}
 \Gamma_s^\mu &= ie_s (2P-2k+q)^\mu, \\
 \Gamma_v^{\mu,\alpha\beta} &=  -i e_v [(2P-2k+q)^\mu g^{\alpha\beta}-(P-k+q)^{\alpha}g^{\mu\beta}\nonumber\\
 &-(P-k)^\beta g^{\mu\alpha}]\label{Gamma},
\end{align}
where $e_{s/v}$ denotes the charge of the scalar/axial-vector diquark.
We point out that, although in the calculation we initially apply Abelian gluons which is very similar to the case of photons, here $e_q$ and $e_X$ are not necessarily to be the electric charges of the quark and diquark.
In fact they can be viewed as the ``color charges" which satisfy the relation $e_{s/v}=- e_q$~\cite{Meissner:2007rx}.
This result comes from the fact that the proton is a color neutral particle, therefore in an Abelian theory the charges of the two components (quark and diquark) should have the same size but opposite sign.

In the following we will calculate $h_L$ and $f_L^\perp$ using two different spectator models.
The first model is the one developed in Ref.~\cite{Bacchetta:2008prd}, the second model is adopted in Ref.~\cite{Bacchetta:plb578}.
There are two differences between these two models, one is the option for the propagator of the axial-vector diquark, the other is the relation between the quark flavors under study and the diquark types.
The two models have also been applied to calculate~\cite{wjmao:2012prd,wjmao:2013epjc} the TMD distributions $g^\perp$ and $e$, which are crucial in the understanding of the beam spin asymmetry measured~\cite{clas04,hermes07,Aghasyan:2011ha,Gohn:2014zbz} in SIDIS.

\begin{figure*}
  \includegraphics[width=0.9\columnwidth]{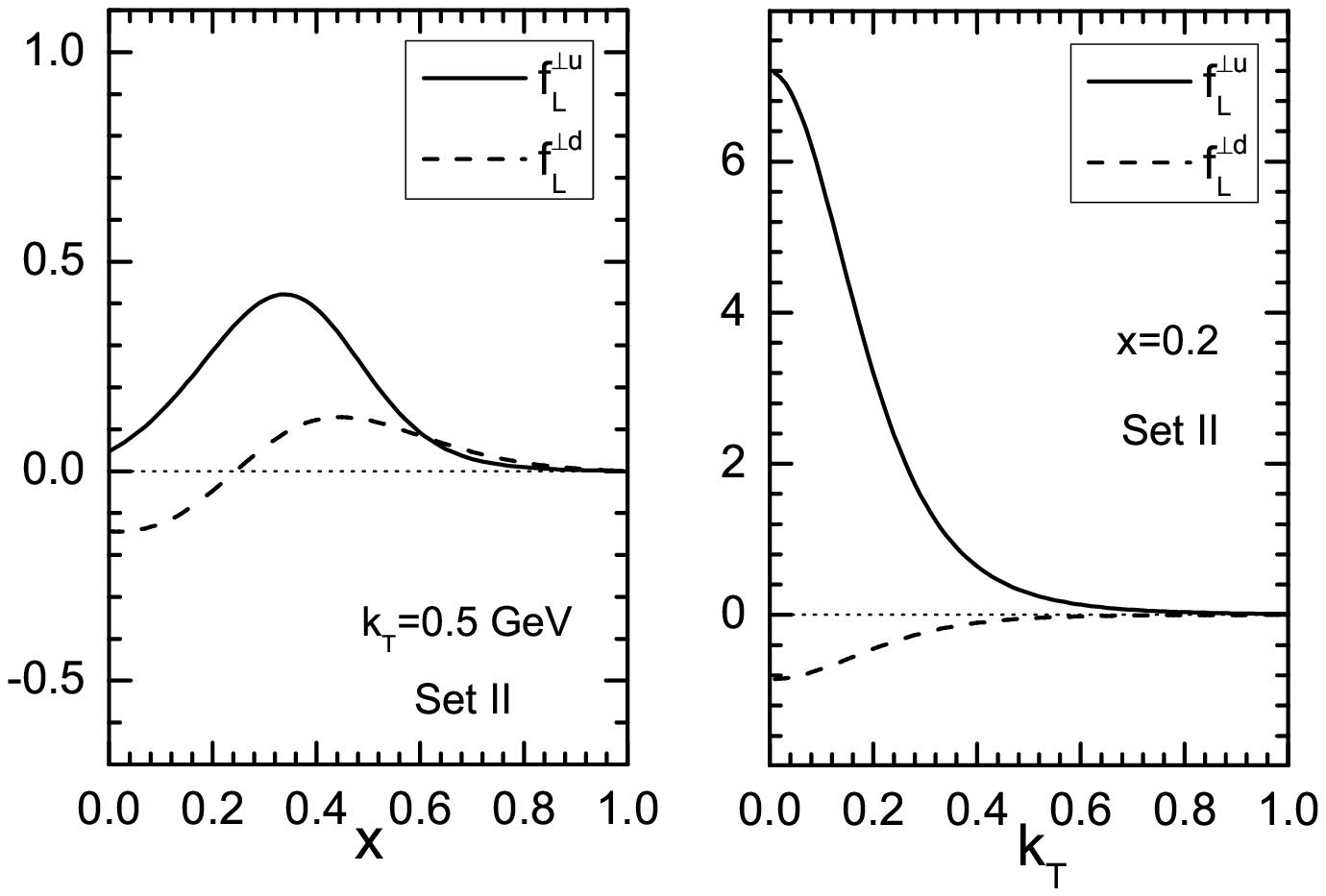}~~~~~~
    \includegraphics[width=0.9\columnwidth]{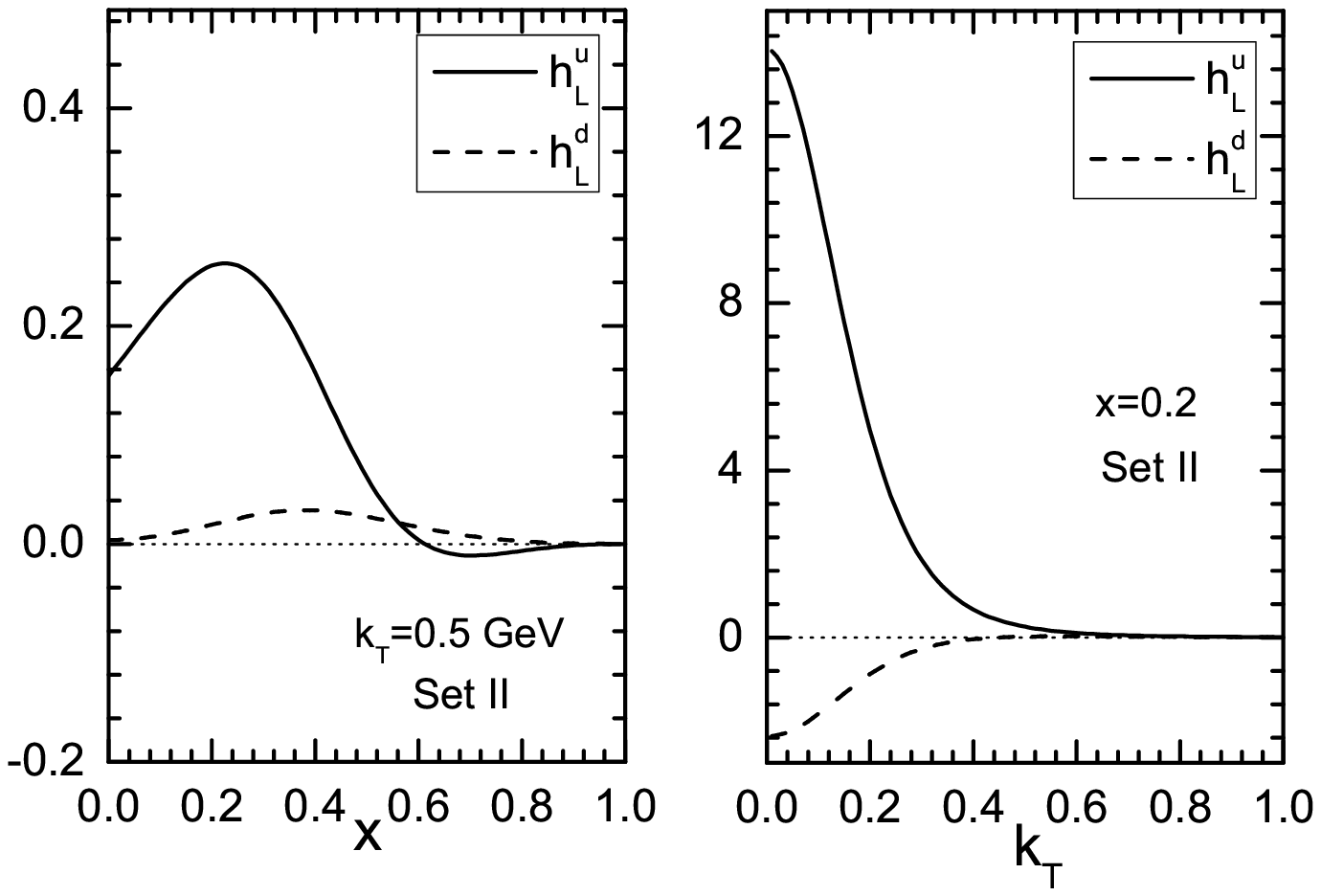}
 \caption{The TMD distributions $f_L^\perp (x,\bm k_T^2)$ (left panel)  and $h_L(x,\bm k_T^2)$ (right panel) for $u$ and $d$ quarks in Set II, calculated from the spectator model in Ref.~\cite{Bacchetta:plb578}.}
 \label{fig:hlflset2}
\end{figure*}

Substituting (\ref{lophis}) into (\ref{eq:hltr}) and (\ref{phis1}) into (\ref{phitr2}), we obtain the contributions from the scalar diquark to $h_L$ and $f_L^\perp$:
\begin{align}
h_L^{s}(x,\bm k_T^2)&=\frac{1}{16\pi^3} \frac{N_s^2(1-x)^2}{(\bk^2+L_s^2)^4}\nonumber\\
& \times \left[(1-x)(xM+m)(M+m)\right.\nonumber\\
&\left.+\left(1-2x-{m\over M}\right)\bk^2
-(x+\frac{m}{M})M_s^2\right],
 \label{hls} \\
f_L^{\perp\,s}
  (x,k_T^{\,2})&=-\frac{N_s^2(1-x)^2}{32\pi^3}{e_s e_q\over4\pi}{1\over L (L+\vec k_T^{\,2})^3}\nonumber\\
&\times  \left[(1-x)((1+2x)M^2 +2mM -\Lambda_s^2)\right.
\nonumber\\
&\left.-(1+x)M_s^2\right],\label{fls}
\end{align}
and the expressions are the same in the two models.

In the first spectator model~\cite{Bacchetta:2008prd} we apply here, the propagator $d_{\mu\nu}$ is chosen as
\begin{align}
 d_{\mu\nu}(P-k)  =& \,-g_{\mu\nu}\,+\, {(P-k)_\mu n_{-\nu}
 \,+ \,(P-k)_\nu n_{-\mu}\over(P-k)\cdot n_-}\,\nonumber\\
 & - \,{M_v^2 \over\left[(P-k)\cdot n_-\right]^2 }\,n_{-\mu} n_{-\nu} ,\label{d1}
\end{align}
which is the summation over the light-cone transverse polarizations of the axial-vector diquark~\cite{Brodsky:2000ii}, and has been applied to calculate the leading-twist TMD distributions in Ref.~\cite{Bacchetta:2008prd}.
With the propagator (\ref{d1}), we arrive at the following expressions for $h_L$ and $f_L^\perp$ from the axial-vector diquark component in the first model:
\begin{align}
h_L^v(x,\bk^2)
   &=\frac{N_v^2(1-x)^2}{16\pi^3}
   \frac{1}{(\bk^2+L_v^2)^4}\,\nonumber\\
   &\times \left[\left(x +{m\over M}\right)\left(M_v^2-M^2(1-x)\right)-m^2(1-x)\right.\nonumber\\
   &\left.-\bm k_T^2 \left({1+x\over 1-x}{m\over M}-1\right)\right],\label{hv1}\\
f_L^{\perp\,v}(x,\bm k_{T}^2)
  &= \frac{N_v^2(1-x)^2}{32\pi^3}{e_v e_q\over4\pi}
  \biggl\{\frac{1}{L_v^2(\bm k_T^2+L_v^2)^3}\,\nonumber\\
&\times\left[\left(1+x-2x^2\right)M^2 +2(1-x)mM \right.\nonumber\\
& \left.-m^2-(1+x)M_v^2+x\Lambda_v^2\right]\nonumber\\
&-\frac{x}{(1-x)(\bm k_T^2+L_v^2)^2\bm k_T^2}\ln\left({\bm k_T^2+L_v^2\over L_v^2}\right)\biggr\}.
\label{fv1}
\end{align}

Also, in this model, a general relation between quark flavors and diquark types is adopted~\cite{Bacchetta:2008prd}:
\begin{align}
f^u=c_s^2 f^s + c_a^2 f^a,~~~~f^d=c_{a^\prime}^2 f^{a^\prime}\label{ud},
\end{align}
where $a$ and $a^\prime$ denote the vector isoscalar diquark $a(ud)$ and the vector isovector diquark $a(uu)$, respectively, and
$c_s$, $c_a$ and $c_{a^\prime}$ are the parameters of the model.
In Ref.~\cite{Bacchetta:2008prd}, these parameters as well as the mass parameters (such as the diquark masses $M_X$, cut-off parameters $\Lambda_X$) are fitted from the ZEUS~\cite{zeus} and GRSV01~\cite{grsv01} distribution sets.
Particularly, the mass parameters for different vector diquark types are treated differently, that is, the two isospin states of the vector diquark are distinguished.
Finally, to connect the ``Abelian" version of the gluon interaction to the QCD color interaction, we apply the replacement~\cite{Brodsky:2002plb}
\begin{align}
|e_q e_X| \rightarrow 4\pi C_F \alpha_S,
\end{align}
and we choose $C_F=4/3$ and $\alpha_S\approx 0.3$, following the choice in Refs.~\cite{Bacchetta:2008prd,Bacchetta:plb578}.
Then we obtain the numerical results of $h_L$ and $f_L^\perp$ and label them as the Set I TMD distributions.
In Fig.~\ref{fig:hlflset1} we plot the $x$-dependence (at $k_T=0.5 ~\textrm{GeV}$) and $k_T$-dependence (at $x=0.2$) of the functions $h_L^q(x,\bm k_T^2)$ and $f_L^{\perp q}(x,\bm k_T^2)$ for $q= u$ and $d$ in Set I.

The second form for the propagator of the vector diquark is
\begin{align}
d_{\mu\nu}(P-k)  =& \,-g_{\mu\nu}.
\label{d2}
\end{align}
which was applied in Ref.~\cite{Bacchetta:plb578}.
Then we obtain the alternative expressions for $h_L^{v}$ and $f_L^{\perp v}$:
\begin{align}
h_L^v(x,\bk^2)
   &=\frac{N_v^2(1-x)^2}{16\pi^3}
   \frac{1}{M(\bk^2+L_v^2)^4}\,\nonumber\\
   &\times \left[m \left(\bm k_{T}^2+M^2 \left(x^2-1\right)+M_v^2\right)\right],\label{hv2}\\
f_L^{\perp \,v}(x,\bm k_{T}^2)
  &= \frac{N_v^2(1-x)}{64\pi^3}{e_v e_q\over4\pi}
  \biggl\{\frac{(1-x)}{L_v^2(\bm k_T^2+L_v^2)^3}\,\nonumber\\
&\times\left[(2+2x-3x^2)M^2 +2mM(1-x)-m^2 \right.\nonumber\\
& \left.-(1-x)\Lambda_v^2-{x\over(1-x)}\bm k_T^2-{2+2x-x^2\over 1-x}M_v^2\right]\nonumber\\
&-\frac{x}{(\bm k_T^2+L_v^2)^2\bm k_T^2}\ln\left({\bm k_T^2+L_v^2\over L_v^2}\right)\biggr\}.
\label{fv2}
\end{align}
which are different from the forms obtained in the first model.

Different from Eq.~(\ref{ud}), another way that has been applied to construct distributions of the $u$ and $d$ valence quarks can be expressed as follows~\cite{Bacchetta:plb578,Jakob:1997npa} :
\begin{align}
f^u=\frac{3}{2}f^s+\frac{1}{2} f^a,~~~~f^d=f^{a^\prime}, \label{set2}
\end{align}
here the coefficients $3/2$, $1/2$ and $1$ in front of $f^X$s are obtained from the SU(4) spin-flavor symmetry of the proton wave function.
In this case, the mass parameters for different axial diquark are the same. We use the relation in (\ref{set2}) together with the expressions (\ref{hls}), (\ref{fls}), (\ref{hv2}) and (\ref{fv2}) to obtain another set of TMD distributions (labeled as Set II).
In this calculation we apply the values for the parameters from Ref.~\cite{Bacchetta:plb578}.
In Fig.~\ref{fig:hlflset2} we plot the $x$ dependence (at $k_T=0.5~\textrm{GeV}$) and $k_T$ dependence (at $x=0.2$) of $h_L^q(x,\bm k_T^2)$ and $f_L^{\perp q}(x,\bm k_T^2)$ for $q= u$ and $d$ in Set II.
Comparing Fig.~\ref{fig:hlflset1} with Fig.~\ref{fig:hlflset2}, we can see that the TMD distributions in Set I are very different from those in Set II.
In Set I the TMD distributions peak at the small $x$ region, while in Set I they peak at the moderate $x$ region.
The $k_T$ dependence of the TMD distributions shows that the $k_T$ widths for $h_L^q(x,\bm k_T^2)$ and $f_L^{\perp q}(x,\bm k_T^2)$  in Set I is larger than those in Set II,  as the sizes of the TMD distributions in Set II decrease very quickly with increasing $k_T$.

\section{Numerical results for LTSA at HERMES, CLAS and COMPASS}
\label{BSAs}

\begin{figure}
  \includegraphics[width=0.9\columnwidth]{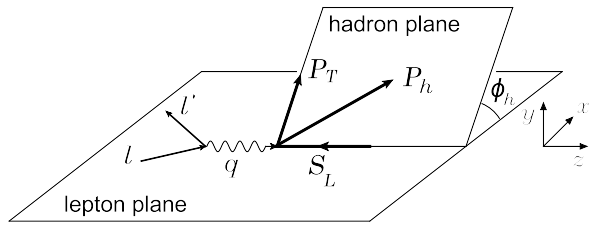}
 \caption{The kinematical configuration for the SIDIS process. The initial and scattered leptonic momenta define the lepton plane ($x-z$ plane), while the detected hadron momentum together with the $z$ axis identify the hadron production plane, the longitudinal spin of the nucleon is along the $-z$ axis.}
 \label{SIDISframe}
\end{figure}

In this section, we will perform the phenomenological analysis
on the LTSA for pions in SIDIS
scattering on a longitudinally polarized nucleon target:
\begin{align}
l (\ell) \, + \, p^\rightarrow (P) \, \rightarrow \, l' (\ell')
\, + \, \pi (P_h) \, + \, X (P_X)\,,
\label{sidis}
\end{align}
at the kinematics of HERMES, CLAS and COMPASS.
The kinematics of SIDIS can be expressed by the following invariant variables:
\begin{align}
&x = \frac{Q^2}{2\,P\cdot q},~~~
y = \frac{P \cdot q}{P \cdot l},~~~
z = \frac{P \cdot P_h}{P\cdot q},~~~\gamma={2Mx\over Q},~~~\nonumber\\
&Q^2=-q^2, ~~~
s=(P+\ell)^2,~~~
W^2=(P+q)^2,~~~
\end{align}
here $q=\ell-\ell'$ is the momentum of the virtual photon, and $W$ is the invariant mass of the hadronic final state.
The reference frame we adopt in this work is shown in Fig.~\ref{SIDISframe}, where the momentum of the virtual photon is along the $z$ axis.
The longitudinal polarization of the target is along the $-z$ axis.
In this frame the transverse momentum of pion is denoted by $\bP$, while the azimuthal angle of the pion around the momentum of the virtual photon is defined as $\ph$.

\begin{figure*}
  \includegraphics[width=0.67\columnwidth]{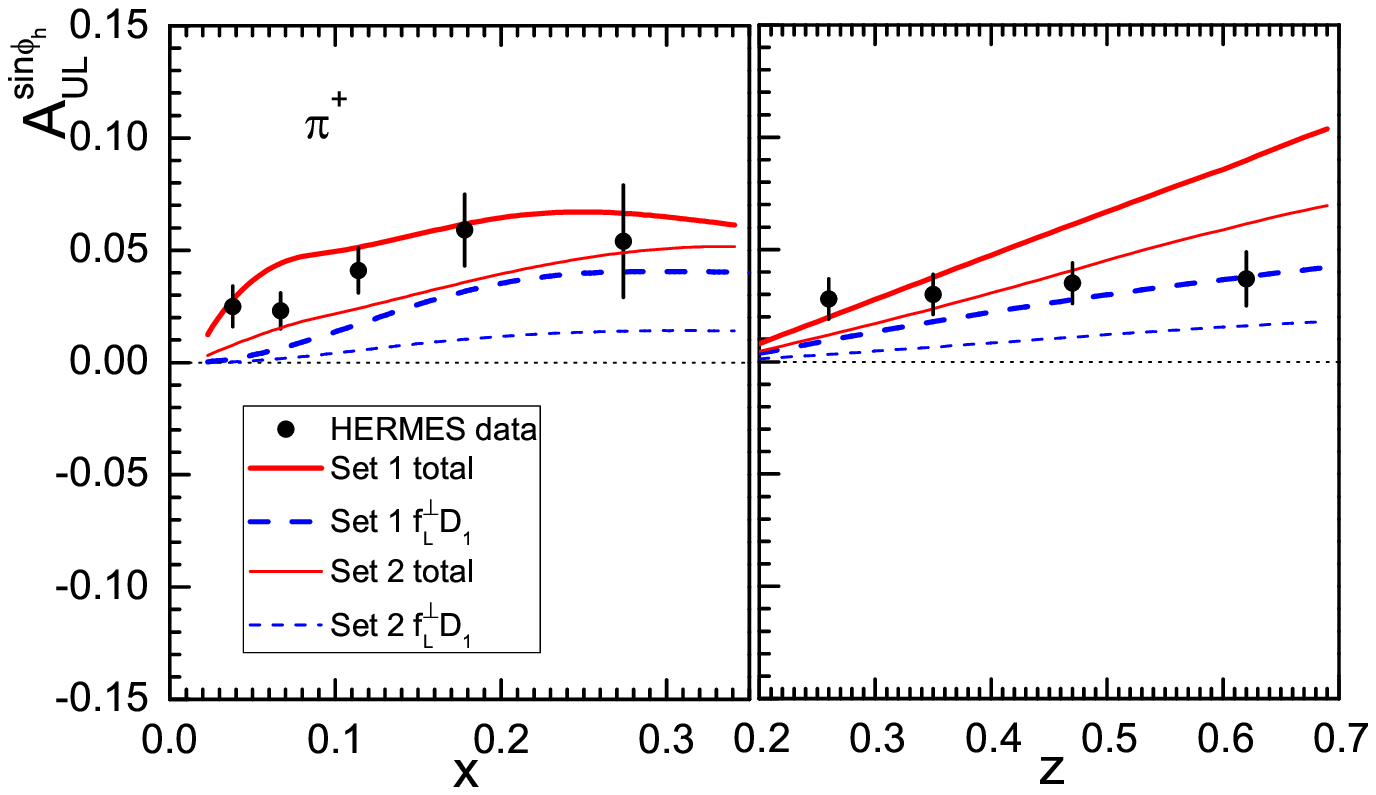}
  \includegraphics[width=0.67\columnwidth]{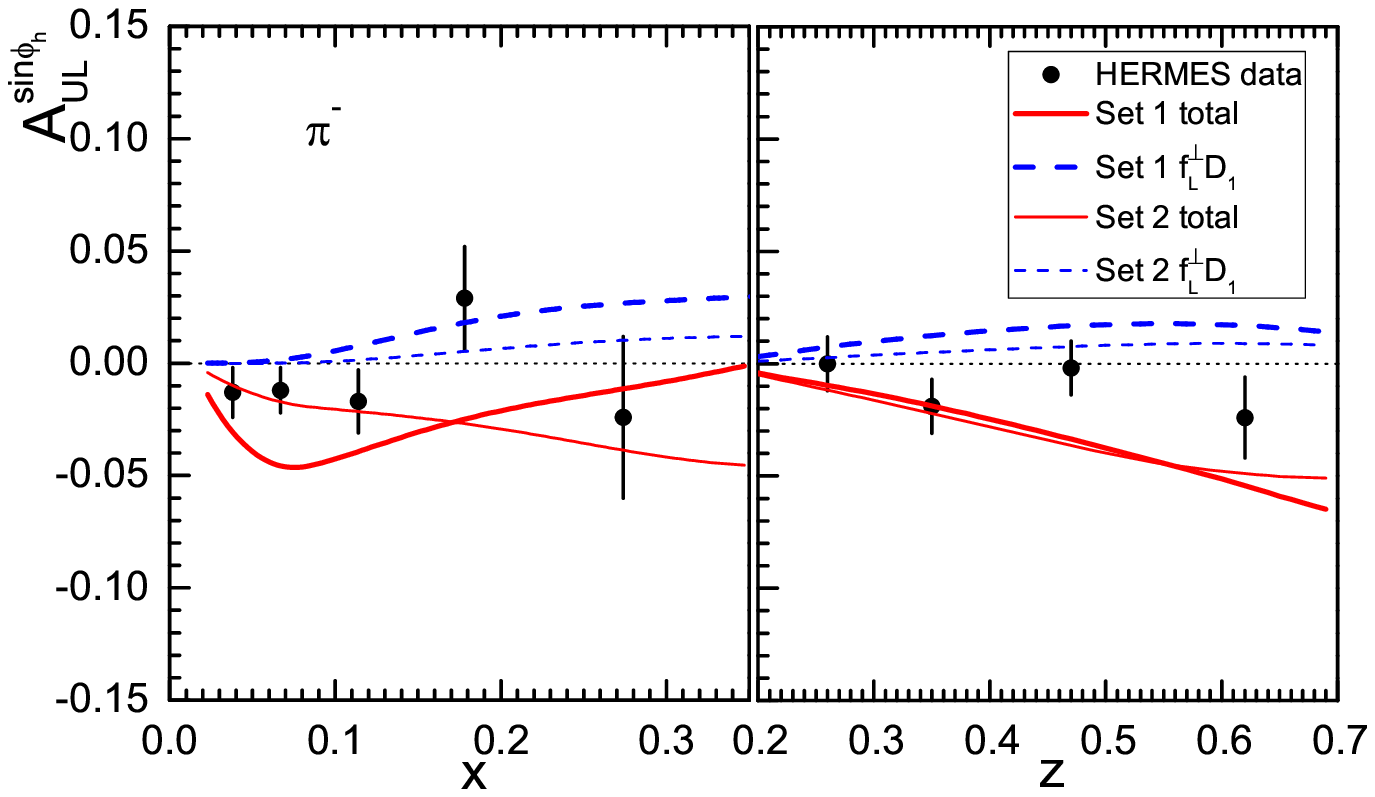}
  \includegraphics[width=0.67\columnwidth]{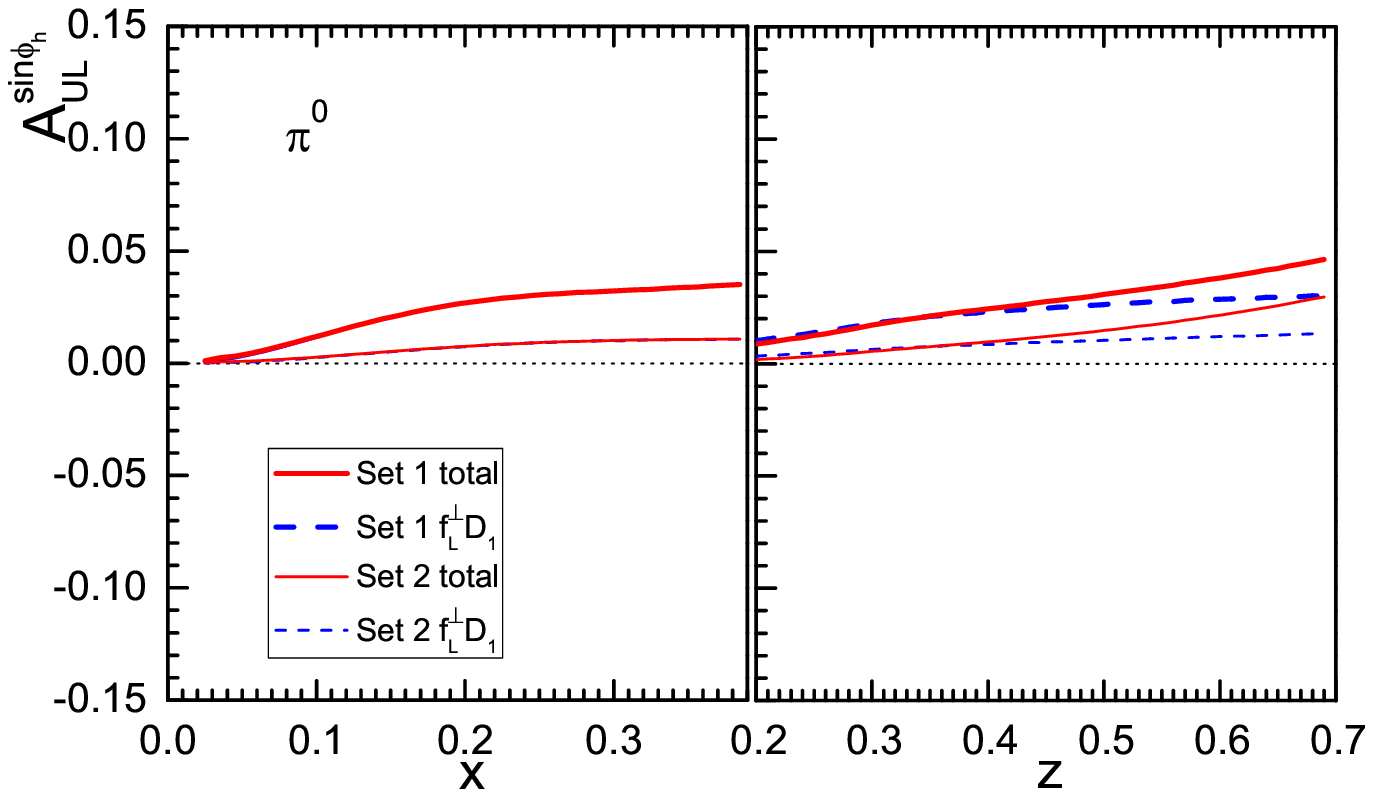}
\caption{The asymmetry $A_{UL}^{\sin\phi_h}$ at HERMES for $\pi^+$
(left panel), $\pi^-$ (central panel) and $\pi^0$ (right panel) as functions of $x$ and $z$. The thick lines and the thin lines are calculated from the Set 1 and Set 2 TMD distributions, respectively. The solid lines represent the total asymmetries, while the dashed lines shows the asymmetries from the $f_L^\perp D_1$ term, the difference between them gives the asymmetries from the $h_L H_1^\perp$ term. Data are from
Ref.~\cite{Airapetian:2005jc}.
}
\label{fig:hermes}
\end{figure*}

The differential cross section of SIDIS scattering on a longitudinally polarized target can be expressed as~\cite{Bacchetta:0611265}:
\begin{align}
\frac{d\sigma}{d\xbj dy\,d\zh dP^2_T d\ph} &=\frac{2\pi \alpha^2}{\xbj y Q^2}\frac{y^2}{2(1-\varepsilon)}
 \Bigl( 1+ \frac{\gamma^2}{2\xbj} \Bigr)
  \left\{ F_{UU} \right.\nonumber\\
  & + \left. S_{\parallel} \sqrt{2\varepsilon(1+\varepsilon)} \sin \phi_h \,\,F^{\sin \ph}_{UL}+\cdots \right\},\label{eq:cs}
\end{align}
where $F_{UU}$ and $F_{UL}^{\sin\phi_h}$ are the spin-averaged and spin-dependent structure functions, respectively.
The ellipsis stands for the leading-twist contribution to the $\sin2\phi_h$ moment, which will not be considered in this paper.
The ratio of the longitudinal and transverse photon flux is given as:
\begin{align}
\varepsilon=\frac{1-y-\gamma^2y^2/4}{1-y+y^2/2+\gamma^2y^2/4}.
\end{align}

In the parton model, based on the tree-level factorization adopted in Ref.~\cite{Bacchetta:0611265}, the two structure functions in Eq.~(\ref{eq:cs}) can be expressed as the convolutions of twist-2 and twist-3 TMD distribution and fragmentation functions.
With the help of the notation
\begin{align}
\mathcal{C}[w fD] &=x\sum_q e_q^2\int d^2\bm k_T\int d^2 \bm p_T\delta^2(z\bm k_T-\bm P_T+\bm p_T) \nonumber\\
&\times w(\bm k_T, \bm p_T)f^q(x,\bm k_T^2) D^q(z,\bm p_T^2),
\end{align}
and the reference frame we choose, $F_{UU}$ and $F_{UL}^{\sin\phi_h}$ have the following forms~\cite{Bacchetta:0611265}
\begin{align}
F_{UU} & = \mathcal{C}[f_1 D_1], \label{eq:FUU}\\
F^{\sin \ph}_{UL} & =  \frac{2M}{Q} \,
\mathcal{C}\,
   \left[\frac{\boldsymbol{\hat{P}_{T}} \cdot \boldsymbol{p_T}}{z M_h}
         \left( \xbj\, h_L H_1^{\perp} + \frac{M_h}{M}\,g_{1L}\, \frac{\tilde{G^{\perp}}}{z} \right)\right.\nonumber\\
&\left.+\frac{\boldsymbol{\hat{P}_{T}}\cdot
    \boldsymbol{k_T}}{M}\left(\xbj\, f_L^{\perp} D_1-\frac{M_h}{M}\, h_{1L}^{\perp} \frac{\tilde{H}}{z} \right)\right] ,\label{eq:FUL}
\end{align}
where $\hat {\bm P}_T= {\bP\over P_T}$ with $P_T =|\bP|$, and $M_h$ is the mass of the final-state hadron.

The asymmetry $A_{UL}^{\sin\phi}$ as a function of $P_T$ therefore can be expressed as
\begin{align}
&A_{UL}^{\sin\phi_h}(P_T) \nonumber\\
&= \frac{\int dx \int dy \int dz \;\frac{1}{x y Q^2}\frac{y^2}{2(1-\varepsilon)}\Bigl( 1+ \frac{\gamma^2}{2x} \Bigr) \sqrt{2\varepsilon(1+\varepsilon)} \;F_{UL}^{\sin\phi_h}}
{\int dx \int dy \int dz \;\frac{1}{x y Q^2}\frac{y^2}{2(1-\varepsilon)}\Bigl( 1+ \frac{\gamma^2}{2x} \Bigr) \;F_{UU}}. \label{eq:asy}
\end{align}
The $x$-dependent and the $z$-dependent asymmetries can be defined in a similar way.

To obtain Eq.~\ref{eq:FUL}, we have assumed that the TMD factorization can be generalized to the twist-3 level.
However, When dealing with higher-twist observables in the TMD framework, one should keep in mind that it is not at all clear from a theoretical point of view if TMD factorization is valid.
In fact, even the tree-level parton model results at twist-3 were questioned, as shown in \cite{Gamberg:2006ru,Bacchetta:2008xw}.
Nevertheless,  we would like to adopt a more phenomenological way, i.e., to use the tree level result in Ref.~\cite{Bacchetta:0611265} to perform the estimate, since there is no alternative theoretical approach for the LTSA so far.

Eq.~(\ref{eq:FUL}) shows that $F^{\sin \ph}_{LU}$ receives four contributions from the convolutions of the twist-3 TMD distribution and fragmentation functions with the twist-2 ones.
In the following calculation, we will apply the Wandzura-Wilczek approximation~\cite{Wandzura:1977qf} to
neglect the contributions from the twist-3 T-odd fragmentation functions $\tilde{G}$ and $\tilde{H}$,
that is, we assume that the sizes of the functions with a tilde are small.
Therefore in our analysis we restrict the scope on the $f_L^\perp D_1$ term and the $h_L H_1^\perp$ term.

\begin{figure*}
  \includegraphics[width=0.68\columnwidth]{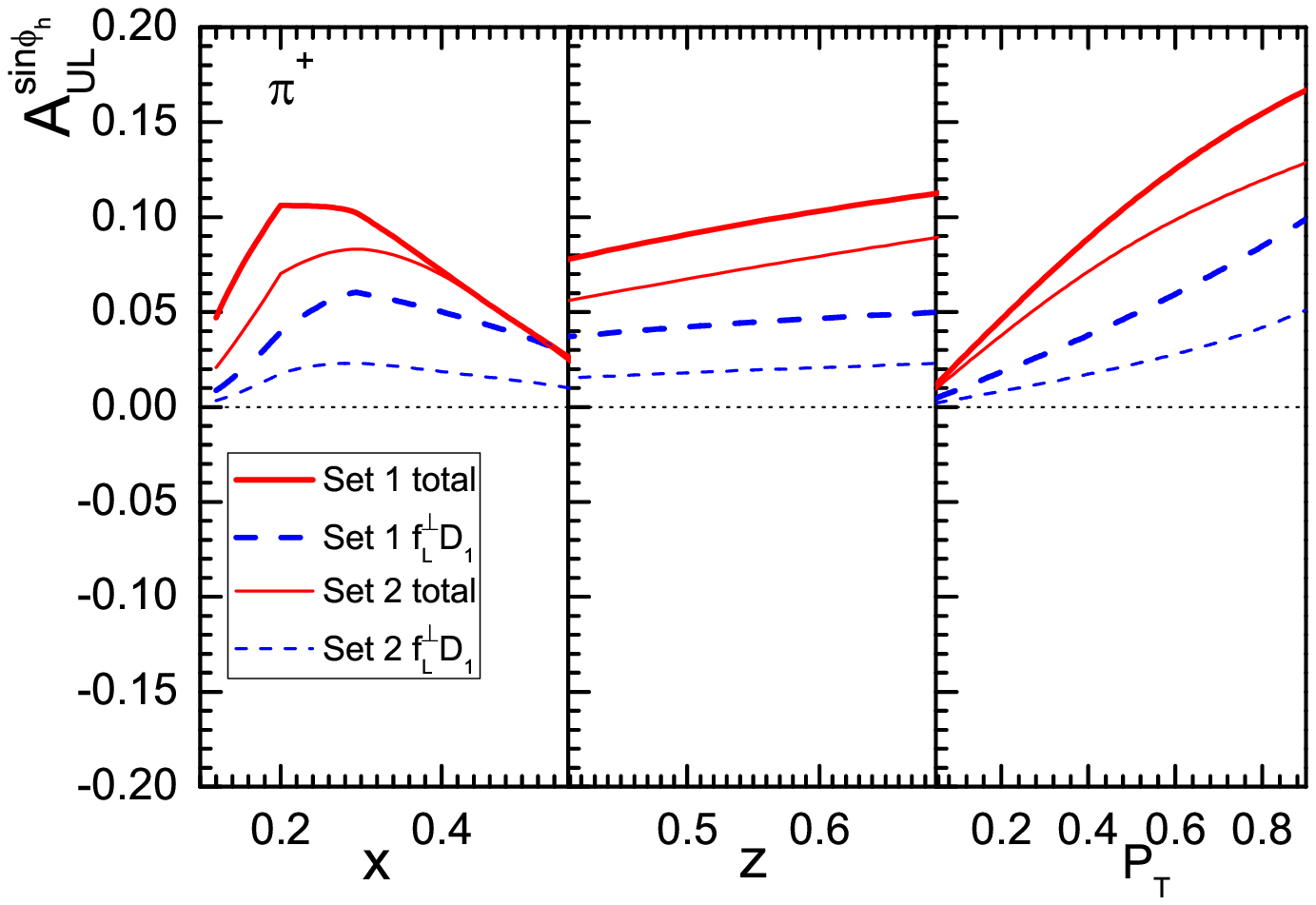}~
  \includegraphics[width=0.68\columnwidth]{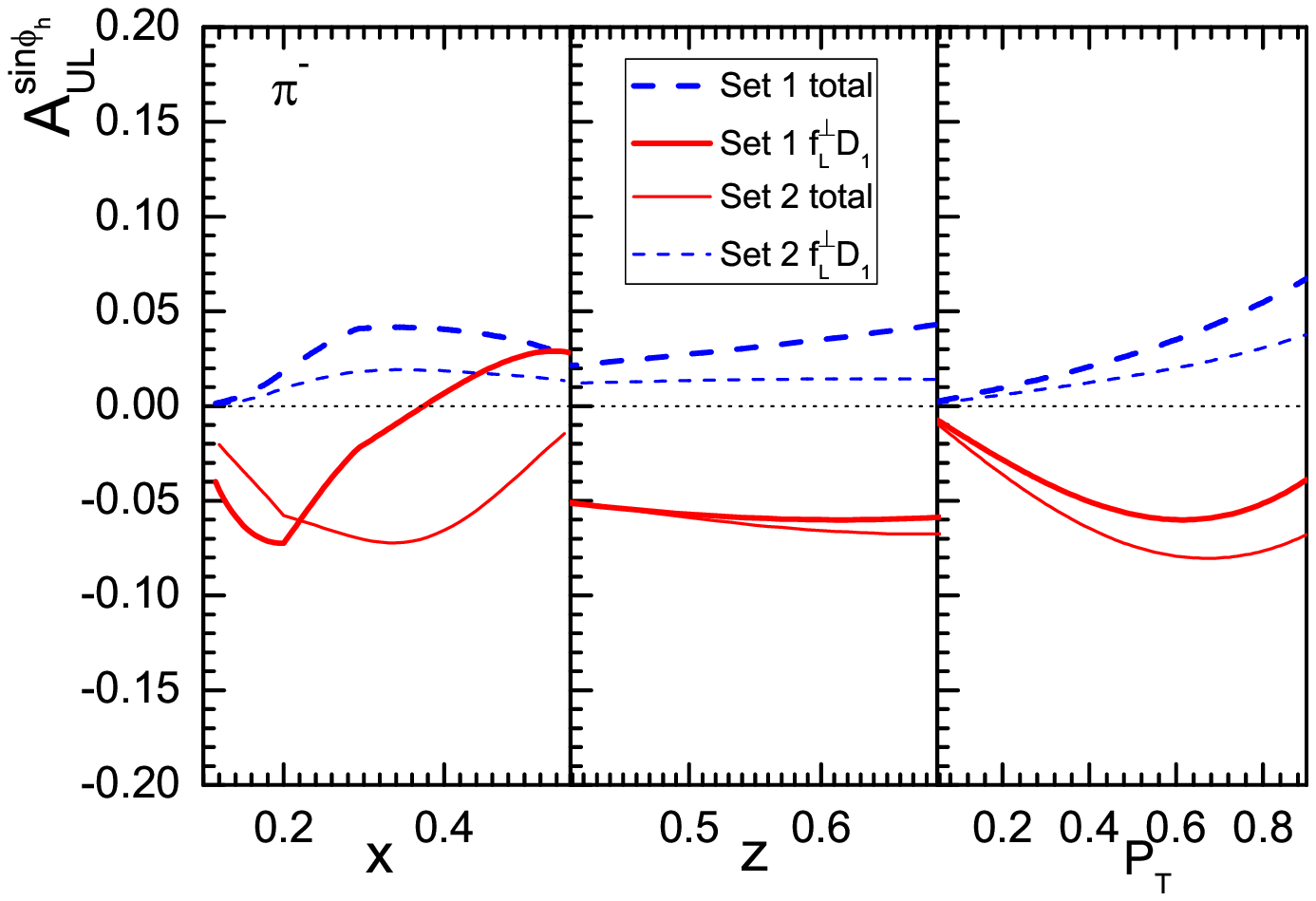}~
  \includegraphics[width=0.69\columnwidth]{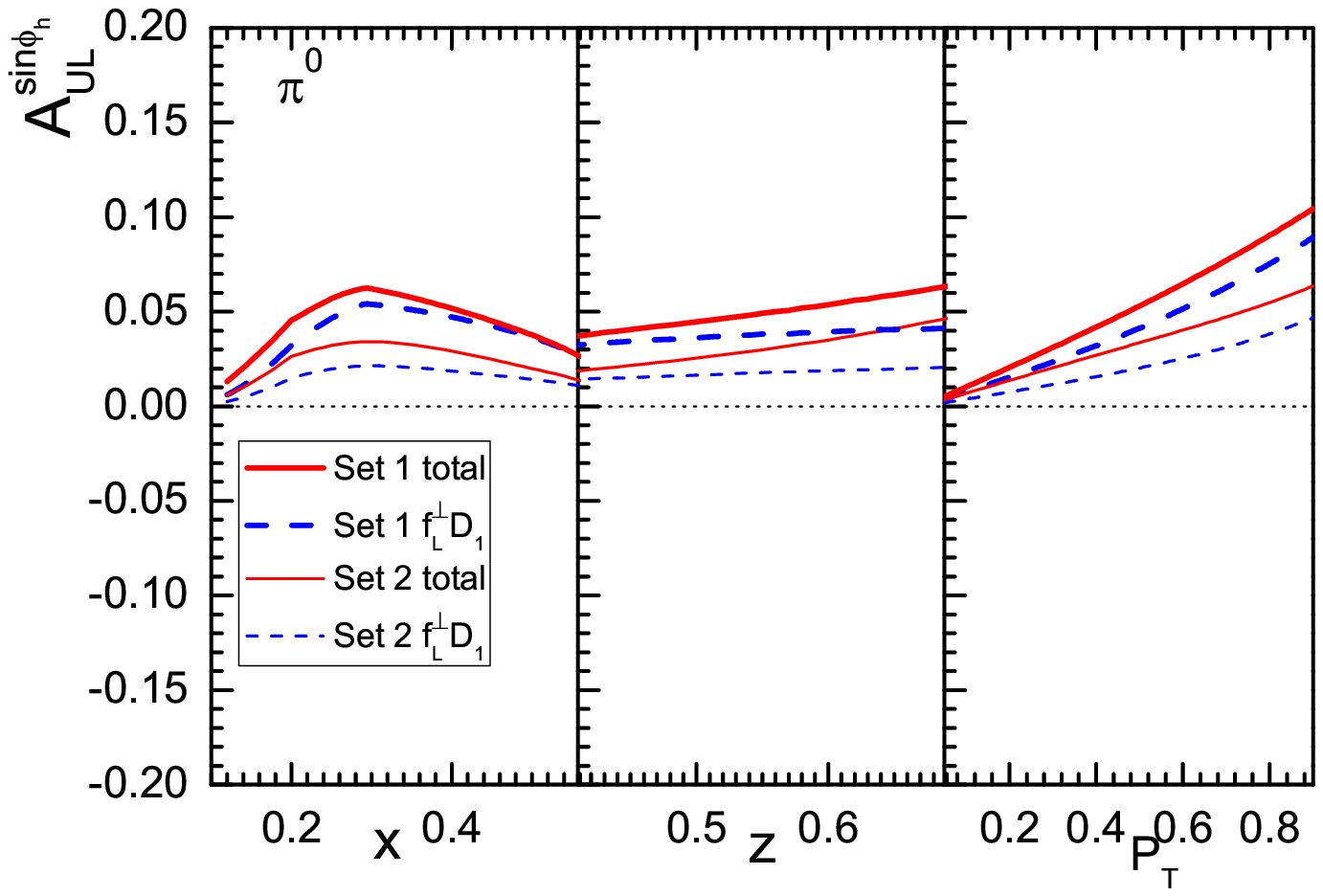}
 \caption{The asymmetry $A_{UL}^{\sin\phi_h}$ for $\pi^+$, $\pi^-$ and $\pi^0$  vs $x$, $z$ and $P_T$ at CLAS 5.5 GeV.}
 \label{fig:clas}
\end{figure*}

For the twist-3 TMD distrbutions $f_L^\perp$ and $h_L$, we apply the results obtained in the previous section.
As for the Collins functions for different pions, we adopt the following relations:
\begin{align}
 H_1^{\perp \pi^+/u}&=H_1^{\perp \pi^-/d}\equiv H_{1 fav}^{\perp} ,\\
 H_1^{\perp \pi^+/d}&=H_1^{\perp \pi^-/u}\equiv H_{1 unf}^{\perp} ,\\
 H_1^{\perp \pi^0/u}&=H_1^{\perp \pi^0/d}\equiv{1\over 2}\left( H_{1 fav}^{\perp}+H_{1 unf}^{\perp}\right),
\end{align}
where $H_{1 fav}^{\perp}$ and $H_{1 unf}^{\perp}$ are the favored and unfavored Collins functions, for which we use the standard parameterisation set from Ref.~\cite{Anselmino:2013vqa}.

For the TMD fragmentation function $D_1^q\left(z,\bp^2\right)$ that couples with the distribution $f_L^\perp$, we assume its $p_T$ dependence  has a Gaussian form
\begin{align}
D_1^q\left(z,\bp^2\right)=D_1^q(z)\, \frac{1}{\pi \langle p_T^2\rangle}
\, e^{-\bm p_T^2/\langle p_T^2\rangle},
\end{align}
where $\langle p_T^2\rangle$ is the Gaussian width for $p_T^2$.
Following the fitted result in Ref.~\cite{Anselmino:2005prd}, we choose $\langle p_T^2\rangle=0.2$ \textrm{GeV}$^2$ in the calculation.
For the integrated fragmentation function $D_1^q(z)$, we will adopt the leading order set of the DSS parametrization~\cite{Florian:2007prd}.
Finally, throughout the paper, we consider the following kinematical constraints~\cite{Boglione:2011} on the intrinsic transverse momentum of the initial quarks in our calculation:
\begin{equation}
 \begin{cases}
k_{T}^2\leq(2-x)(1-x)Q^2, ~~~\textrm{for}~~0< x< 1 
; \\
k_{T}^2\leq \frac{x(1-x)} {(1-2x)^2}\, Q^2, ~~~~~~~~~~~~\textrm{for}~~x< 0.5.
\end{cases}\label{constraints}
 \end{equation}
The first constraint in Eq.~(\ref{constraints}) is obtained by requiring the energy of the parton to be less than the energy of the parent hadron, while the second constraint
arises from the requirement that
the parton should move in the forward direction with respect to the parent hadron~\cite{Boglione:2011}.
For the region $x<0.5$, there are two upper limits for $k_T^2$ applied in the region $x<0.5$ at the same time; it is understood that the smaller one should be chosen.

We adopt the following kinematics to calculate numerically the SSAs $A_{UL}^{\sin\phi_h}$ at HERMES~\cite{Airapetian:2005jc}:
\begin{align}
&0.023 < x < 0.4,\,0.1 < y < 0.85,\, 0.2<z<0.7 \nonumber\\
& W^2 > 10\, \textrm{GeV}^2, \, Q^2 >\, 1 \textrm{GeV}^2,\,4\,\textrm{GeV} < P_{\pi} < 13.8\,\textrm{GeV},\nonumber
\end{align}
where $P_{\pi}$ is the momentum of the final-state pion.
In the left, central, and right panels of Fig.~\ref{fig:hermes}, we show the results of the LTSAs for $\pi^+$, $\pi^-$ and $\pi^0$ as functions of $x$ and $z$.
The thick and thin curves correspond to the asymmetries calculated from the TMD distributions in Set I and Set II, respectively.
The dashed curves show the asymmetries contributed by the $f_L^\perp D_1$ term, while the solid curves denote the total contribution of the $f_L^\perp D_1$ and $h_L H_1^\perp$ terms.
The difference between the solid curve and the dashed curve gives the asymmetry from the $h_L H_1^\perp$ term.
In the case of $\pi^+$, $\pi^-$ production, we compare our results with the HERMES data~\cite{Airapetian:2005jc}, which have been obtained by subtracting the contribution from the transverse spin component.

The results calculated from both sets of TMD distributions show that the asymmetry for $\pi^+$ is positive, while that for $\pi^-$ is negative, consistent with the HERMES data.
In the case of $\pi^+$ production, the asymmetry calculated from the Set I agree with the data better than from the Set II TMD distributions .
In the case of $\pi^-$ production, the asymmetry contributed by the $f_L^\perp D_1$ term is positive, while the that contributed by the $h_L H_1^\perp$ term is largely negative, leading to a negative asymmetry in total.
In the case of $\pi^0$ production, the asymmetries is positive, similar to the case of $\pi^+$, but the size is smaller.
An important feature in the $\pi^0$ asymmetry is that the contribution from the $f_L^\perp D_1$ term  dominates, and that the one from the $h_L H_1^\perp$ term is almost negligible, especially in the $x$-dependent asymmetry.
This is very different from the case of charged pions, for which the contributions from the $f_L^\perp D_1$ term and the $h_L H_1^\perp$ term are comparable.
Therefore the measurement on the asymmetry $A_{UL}^{\sin\phi_h}$ for $\pi^0$ at HERMES will provide great opportunity to access the distribution $f_L^\perp$.

\begin{figure*}
  \includegraphics[width=0.67\columnwidth]{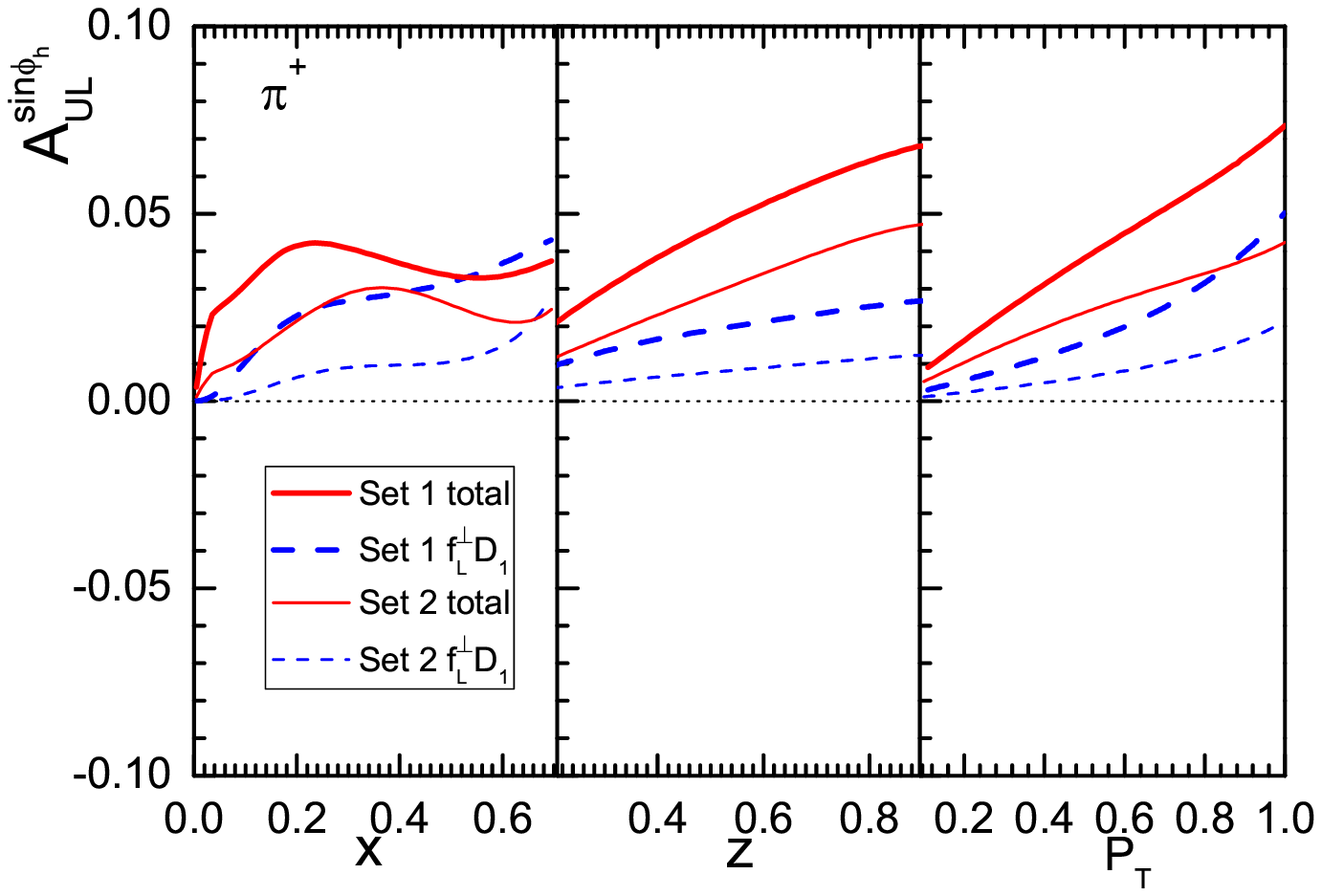}
  \includegraphics[width=0.67\columnwidth]{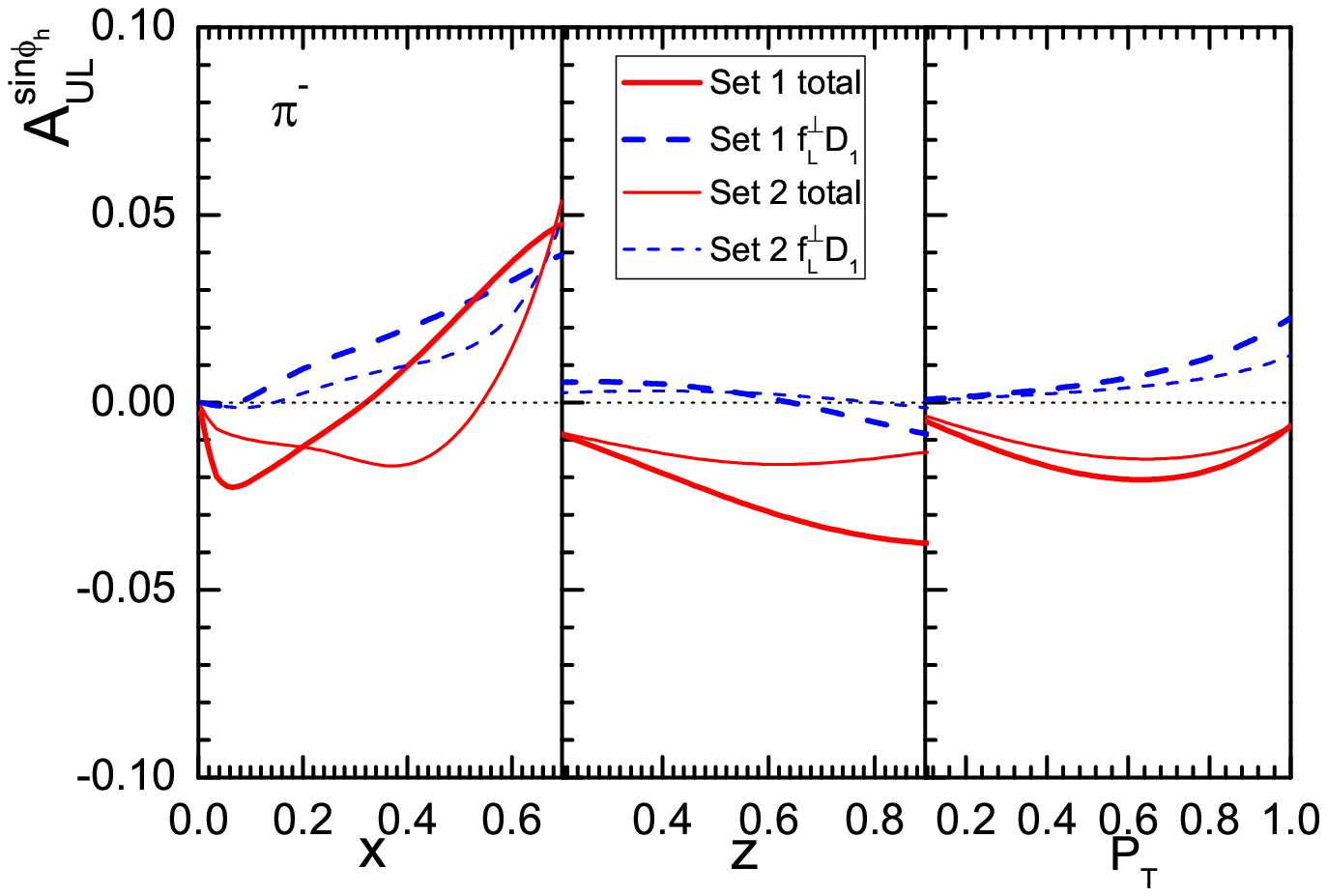}
  \includegraphics[width=0.67\columnwidth]{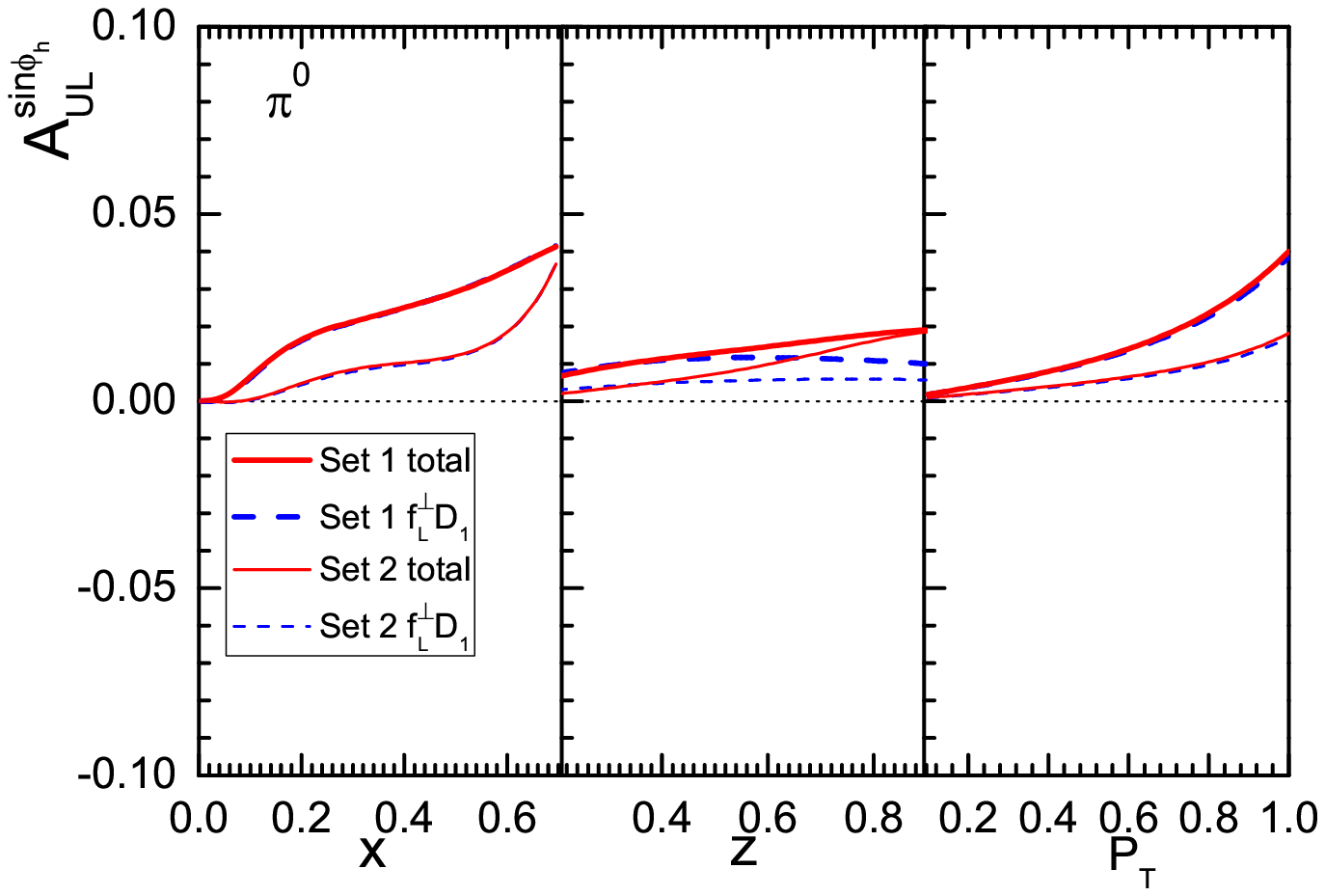}
 \caption{The asymmetry $A_{UL}^{\sin\phi_h}$ for $\pi^+$, $\pi^-$ and $\pi^0$  at COMPASS}
 \label{fig:compass}
\end{figure*}

The asymmetry $A_{UL}^{\sin\phi_h}$ for all three pions is also being measured at CLAS~\cite{Avakian:2013sta} by using an electron beam ($E_e=5.5~\textrm{GeV}$) scattered off the longitudinally polarized proton target.
The kinematical cuts at CLAS applied in our calculation are:
\begin{align}
&0.1<x<0.6,~~ 0.4<z<0.7,~~ Q^2>1\, \textrm{GeV}^2,\nonumber\\
&P_T>0.05\,\textrm{GeV},~~ W^2>4\,\textrm{GeV}^2. \nonumber
\end{align}
In Fig.~\ref{fig:clas}, we show the asymmetry $A_{UL}^{\sin\phi_h}$ for $\pi^+$, $\pi^-$ and $\pi^0$ at CLAS.
For all three pions the asymmetries are sizable and could be measured at CLAS.
For $\pi^+$ and $\pi^0$ production, the asymmetries calculated from Set I TMD distributions are larger than those from Set II distributions; while
for $\pi^-$ production, the size of the asymmetry from Set II TMD distributions is slightly larger than that from Set I.
In the case of $\pi^0$ production, again the $f_L D_1$ term dominates the asymmetries as functions of $x$ and $P_T$; although there is sizable contribution from the $h_L H_1^\perp$ term at the high $z$ region.

COMPASS has measured~\cite{Alekseev:2010dm} the $\sin\phi$ azimuthal asymmetries of charged hadrons produced off the longitudinally polarized deuteron target with a 160 GeV muon beam, where $\phi$ is the azimuthal angle of the final-state hadron around the momentum of the lepton beam.
The azimuthal asymmetry $A_{UL}^{\sin\phi_h}$ at COMPASS might be extracted by the method in Ref.~\cite{Airapetian:2005jc}.
Thus, we calculate the asymmetry $A_{UL}^{\sin\phi_h}$ of charged and neutron pions produced off a deuteron target at COMPASS.
We adopt the following kinematics in the estimation:
\begin{align}
&0.004<x<0.7,~~y>0.1,~~0.2<z<0.9,~~ \nonumber\\
&x_F>0,~~Q^2>1\, \textrm{GeV}^2,~~0.1 \textrm{GeV}<P_T<1\,\textrm{GeV},~~ \nonumber \\
 & 5 \,\textrm{GeV}<W<18\,\textrm{GeV}. \nonumber
\end{align}
In Fig.~\ref{fig:compass}, we plot the asymmetry
$A_{UL}^{\sin\phi_h}$ for $\pi^+$, $\pi^-$ and $\pi^0$ vs $x$, $z$ and $P_T$.
We find that in the case of deuteron target, again, for $\pi^0$ production the dominant contribution comes from the $f_L^\perp D_1$ term, while the contribution from the $h_L H_1^\perp$ term is consistent with zero.
Therefore, it would be feasible to extract the flavor dependence of $f_L^\perp$ by combining the asymmetries for $\pi^0$
using a deuteron target together with a proton target.

\section{conclusion}
\label{conclusion}
In this work, we studied the subleading-twist effect on the azimuthal asymmetry of pions produced in SIDIS with longitudinal polarized nucleon targets.
We considered the contributions from the twist-3 TMD distributions $f_L^\perp$ and $h_L$ simultaneously.
In our calculation, two different sets of twist-3 distribution functions are employed to analyze the asymmetry $A_{UL}^{\sin\phi_h}$ for $\pi^+$, $\pi^-$ and $\pi^0$ at HERMES, CLAS and COMPASS.
By comparing our results with the HERMES measurement in which a proton target is used, we find that
the sign of the asymmetries for $\pi^+$ and $\pi^-$ is consistent with the HERMES data,
and the $f_L^\perp D_1$ and $h_L H_1^\perp$ terms both give substantial contributions to $A_{UL}^{\sin\phi_h}$.
In the case of $\pi^-$ production, the $h_L H_1^\perp$ term is largely negative, leading to the negative sign of the total asymmetry.
We also find that the $f_L^\perp D_1$ term dominates in the asymmetry for $\pi^0$, especially at HERMES and COMPASS where the beam energies are higher.
Thus the measurement of $A_{UL}^{\sin\phi_h}$ for $\pi^0$ on the proton target, as well as on the deuteron target, will provide great opportunity to access $f_L^\perp$, including its flavor separation.
Future data on $A_{UL}^{\sin\phi_h}$ for charged and neutral pions from CLAS at 5.5 GeV may be used to clarify the role of twist-3 TMD distribution functions on the longitudinal target spin asymmetry.

\section*{Acknowledgements}
The author is grateful to H. Avakian and W. Gohn for useful discussions.
This work is partially supported by the National Natural Science
Foundation of China (Grant Nos.~11005018 and~11120101004),  by the Fundamental Research
Funds for Central Universities (Grant No.~2242012R3007), and by the Qing Lan Project.

\end{document}